%% file: main.tex
\documentclass[conference]{IEEEtran}

\input{packages}
\input{macro}

\makeatletter

\IEEEoverridecommandlockouts

\def\BibTeX{{\rm B\kern-.05em{\sc i\kern-.025em b}\kern-.08em
    T\kern-.1667em\lower.7ex\hbox{E}\kern-.125emX}}
\begin{document}

\title{\huge Interference-free Operating System: A 6 Years' Experience \\in Mitigating Cross-Core Interference in Linux} 


\author{\IEEEauthorblockN{Zhaomeng Deng}
\IEEEauthorblockA{\textit{Key Lab of HCST (PKU), MOE;} \\
\textit{SCS, Peking University} \\
Beijing, China \\
infinityedge@pku.edu.cn}
\and
\IEEEauthorblockN{Ziqi Zhang}
\IEEEauthorblockA{\textit{Key Lab of HCST (PKU), MOE;} \\
\textit{SCS, Peking University} \\
Beijing, China \\
ziqi\_zhang@pku.edu.cn}
\and
\IEEEauthorblockN{Ding Li\IEEEauthorrefmark{1}\thanks{* are the corresponding authors.}}
\IEEEauthorblockA{\textit{Key Lab of HCST (PKU), MOE;} \\
\textit{SCS, Peking University} \\
Beijing, China \\
ding\_li@pku.edu.cn}
\and
\IEEEauthorblockN{Yao Guo\IEEEauthorrefmark{1}}
\IEEEauthorblockA{\textit{Key Lab of HCST (PKU), MOE;} \\
\textit{SCS, Peking University} \\
Beijing, China \\
yaoguo@pku.edu.cn}
\and
\IEEEauthorblockN{Yunfeng Ye}
\IEEEauthorblockA{\textit{Huawei Technologies}\\
Beijing, China}
\and
\IEEEauthorblockN{Yuxin Ren}
\IEEEauthorblockA{\textit{Huawei Technologies}\\
Beijing, China}
\and
\IEEEauthorblockN{Ning Jia}
\IEEEauthorblockA{\textit{Huawei Technologies}\\
Beijing, China}
\and
\IEEEauthorblockN{Xinwei Hu}
\IEEEauthorblockA{\textit{Huawei Technologies}\\
Beijing, China}

}
\maketitle

\begin{abstract}
\input{tex/abstract}

\end{abstract}

\IEEEpeerreviewmaketitle

\input{tex/intro}
\input{tex/motiv}

\input{tex/empirical}
\input{tex/lessons}

\input{tex/eval}
\input{tex/conc}
\input{tex/Acknowledgments}
\bibliographystyle{IEEEtran} 
\bibliography{references}

\end{document}

%% file: packages.tex
\usepackage{graphicx}
\usepackage[linesnumbered,lined,vlined,ruled,commentsnumbered,noend]{algorithm2e}
\usepackage{textcomp}
\usepackage{xcolor}
\usepackage{listings}
\usepackage{url}
\usepackage{xspace}
\usepackage{multirow}
\usepackage{balance}
\usepackage{ctable}
\usepackage{listings}
\usepackage{color}
\usepackage{graphicx}
\usepackage{subcaption}
\usepackage[nobiblatex]{xurl}

\usepackage{multirow}

\usepackage{algpseudocode}
\usepackage{caption}
\usepackage{subcaption}

\usepackage{cleveref}
\usepackage{acronym}
\usepackage{fancybox}
\usepackage{verbatim}
\usepackage[normalem]{ulem}
\usepackage{float}
\usepackage{newfloat}
\usepackage{mdframed}
\usepackage{booktabs}
\usepackage{pifont}
\usepackage{paralist}
\usepackage{stfloats}
\usepackage{colortbl}
\usepackage{makecell}

%% file: macro.tex
\definecolor{red}{RGB}{255,0,0}
\definecolor{green}{RGB}{18,220,168}
\definecolor{darkgreen}{RGB}{0,128,0}
\newcommand{\ie}{i.e.,\xspace}
\newcommand{\eg}{e.g.,\xspace}

\newcommand{\etal}{\textit{et al.}\xspace}

\newcommand{\eat}[1]{}
\acrodef{wp}[WP]{Website Fingerprinting}
\acrodef{apt}[APT]{Advanced Persistent Threats}
\acrodef{lol}[LotL]{Living-Off-The-Land}
\acrodef{ids}[IDS]{Intrusion Detection System}
\acrodef{vae}[VAE]{Variational AutoEncoder}
\acrodef{re}[RE]{reconstruction error}
\acrodef{sv}[SV]{Stableness Value}
\acrodef{as}[AS]{anomaly score}
\acrodef{gas}[GAS]{graph anomaly score}
\acrodef{etw}[ETW]{Event Tracing for Windows}
\acrodef{e3}[E3]{Engagement 3}
\acrodef{e5}[E5]{Engagement 5}
\acrodef{ttp}[TTPs]{Tactics, Techniques, and Procedures}
\acrodef{hsg}[HSG]{High-level Scenario Graph}
\acrodef{nlp}[NLP]{Natural Language Processing}
\acrodef{dg}[DG]{Detection Graph}
\acrodef{poi}[POI]{Point of Interest}
\acrodef{iv}[IV]{Important Value}
\acrodef{sg}[SG]{Suspicious Graph}
\acrodef{mttd}[MTTD]{Mean Time to Detect}
\acrodef{soc}[SOC]{Security Operations Center}

\newlength{\MaxSizeOfLineNumbers}%
\definecolor{keywordcolor}{rgb}{0.8,0.1,0.5}
\definecolor{lightlightgray}{gray}{.96}
\definecolor{lightgray}{gray}{.925}
\definecolor{medlightgray}{gray}{0.7}
\definecolor{medgray}{gray}{0.4}
\definecolor{darkgray}{gray}{0.35}
\definecolor{nearblack}{gray}{0.15}

\crefname{component}{Component}{Components}
\newcommand{\head}[1]{\noindent{\bf #1}}

%% file: tex/abstract.tex
Real-time operating systems employ spatial and temporal isolation to guarantee predictability and schedulability of real-time systems on multi-core processors.
Any unbounded and uncontrolled cross-core performance interference poses a significant threat to system time safety.
However, the current Linux kernel has a number of interference issues and represents a primary source of interference.
Unfortunately, existing research does not systematically and deeply explore the cross-core performance interference issue within the OS itself.

This paper presents our industry practice for mitigating cross-core performance interference in Linux over the past 6 years.
We have fixed dozens of interference issues in different Linux subsystems.
Compared to the version without our improvements, our enhancements reduce the worst-case jitter by a factor of 8.7, resulting in a maximum 11.5x improvement over system schedulability.
For the worst-case latency in the Core Flight System and the Robot Operating System 2, we achieve a 1.6x and 1.64x reduction over RT-Linux.
Based on our development experience, we summarize the lessons we learned and offer our suggestions to system developers for systematically eliminating cross-core interference from the following aspects: task management, resource management, and concurrency management.
Most of our modifications have been merged into Linux upstream and released in commercial distributions.

%% file: tex/intro.tex
\section{Introduction}
\label{s:intro}

Given the increased computing capacity and the decrease in size, weight, and power, multi-core processors have been widely used in many real-time systems and latency-critical applications~\cite{survey_rtss_20}, such as data flow~\cite{flow_rtss_20} and object detector in autonomous driving~\cite{object_rtss_20}, mixed-criticality systems~\cite{mcs,separt_vee_14,edfvd_rtss_16}, and robotics~\cite{ros2_rtss_21,rosm_rtss_22}.
Unfortunately, multi-core processors introduce a core challenge of ensuring strong timing guarantees in real-time systems: cross-core performance interference, \ie\ execution on one core causes variable delays to (irrelevant) execution on another core. 
This is not a functional bug because the system's functionality is correct. 
However, the timing behavior of the real-time system is affected. 
In particular, the operating system (OS) that manages all the cores becomes a major source of interference.
The OS isolates a set of cores exclusively for real-time tasks, but uncontrolled cross-core interference breaks spatial and temporal isolation, and fundamentally affects the whole system predictability and schedulability.
We are particularly concerned with the interference introduced by the operating system itself, which impacts strict core isolation.
This interference can lead to delays in task activation, preempt high-priority task execution, and increase task blocking, severely damaging the timing performance of real-time systems.
Our evaluation in \S\ref{s:eval} demonstrates that when interference is considered as task release jitter, it significantly reduces the system's schedulability.

However, completely eliminating cross-core interference from OS is difficult. 
Based on our industry practice in supporting various real-time systems, we point out the two key challenges of mitigating cross-core interference in OS.

\emph{1) Modern OSes compose a large number of complex and intertwined functionalities and services, while lacking a unified view of interference management.}
After years of evolution, commercial OSes 
\footnote{
The OS here refers to the entire execution environment of the system.
It can be either a monolithic kernel, such as Linux, or a micro-kernel~\cite{fasic_rtss_02, sel4_rtss_11} plus various user-level system services.
}
provide rich features with a large code base.
However, these features are often independently developed without unified design guidance about predictability.
Thus, after integrating these functions into one system, there will be many performance interferences that are difficult to detect.

\emph{2) The increasingly complicated platform contains many shared resources without explicit and formal specifications.}
Existing architectures contain massive diverse software and hardware resources.
Complex interaction of different OS functionalities can lead to unintentional resource sharing.
Even worse, these shared resources are not identified and addressed during the design phase, thus they fail to follow the proper sharing protocols and generate unexpected interference throughout various system modules, such as task placement, resource management, and concurrency control.

Existing researches have not systematically explore the cross-core performance interference issue within the OS itself.
On the one hand, many researchers optimize interference around user-level tasks, for example, real-time synchronization protocol~\cite{lock_rtas_08, smr_rtas_18, runner_rtss_21}, task scheduling and partition~\cite{model_rtss_15,part_rtss_16,maracas_rtss_16}, and parallelism management~\cite{fjos_rtas_14,steal_rtss_16}.
Unfortunately, these techniques are not applied directly to the commercial OS due to different execution requirements.
On the other hand, OS researchers deal with an individual source of interference, such as memory~\cite{memory_rtss_19}, cache~\cite{cache_rtas_16}, interrupts~\cite{inter_rtss_08,chaos_rtas_19}, processor affinity~\cite{affinity_rtss_14}, communication~\cite{comm_rtss_14,herd_rtas_22}, and virtualization~\cite{virt_rtas_18}.
However, these technologies are fragmented and independent from each other.
As a result, they cannot guarantee coverage of all interferences, and it is also difficult to integrate them into a single system.
Furthermore, the proposed research prototypes are far from commercial use, and once they grow into mature production-scale systems, they will face the same challenge discussed above.

This paper presents our industry practice for mitigating cross-core performance interference in Linux over the past 6 years.
We support dozens of productions ranging from embedded systems to large multi-core servers.
Avoiding cross-core interference is their paramount requirement for the underlying OS.
Linux is widely used (over 55\%~\cite{survey_rtss_20}) in real-time systems.
While long discussions about creating full performance isolation on Linux were made~\cite{linux_task_iso,timer_lwnet}, and the Linux community strives to avoid interference, due to these challenges, the current Linux implementation contains a large number of bugs causing performance interference.
We have identified and fixed dozens of interference issues in different Linux subsystems.
According to our experience, we argue that discovering and avoiding interference should be systematically considered from the following three aspects: task management, resource management, and concurrency management.
Table \ref{tbl:overview} summarizes our representative work introduced in this paper.

\begin{table*}[t]
  \centering
  \renewcommand\arraystretch{0.7}
  
  \caption{\small Overview of the cross-core interference problems and solutions studied in this paper.}
  \begin{tabular}{cll}
    \toprule
    Category                       & \multicolumn{1}{c}{Problem} & \multicolumn{1}{c}{Solution} \\ \midrule
\multirow{2}{*}{\begin{tabular}[c]{@{}c@{}}Task \\Management\end{tabular}} & Unnecessary worker thread activation in \texttt{workqueue} operations.  & Restricting the activation to the necessary extents.  \\ \cmidrule{2-3}
                             & Incorrect core selection in task migration.    &  Considering the isolated cores in core selection logic.    \\ \midrule 
\multirow{2}{*}{\begin{tabular}[c]{@{}c@{}}Resource \\Management\end{tabular}} & Cross-core interference caused by the exhaustion of ASID.   & Partition
ASID space and integrate ASID into the core isolation.  \\ \cmidrule{2-3}
  & Unnecessary {\tt backlog} flushing in all cores when uninstalling a NIC. &  Add a necessity check, flush only when needed.  \\ \midrule 
\multirow{2}{*}{\begin{tabular}[c]{@{}c@{}}Concurrency \\Management\end{tabular}} &  Concurrent blocking in jiffles synchronization.   & Reducing the critical section length.   \\ \cmidrule{2-3}
  & Statistics system aggregates data from all cores may produce interference. &  Restrict the process of statistics updating to within partitions.    \\ \bottomrule 
  \end{tabular}
  \label{tbl:overview}

\end{table*}

Based on our development and deployment experience, we have learned lessons on further analyzing and eliminating cross-core interference.
Our lessons cover four aspects: 
\begin{inparaenum}[1)]
\item using unified isolation mechanisms,
\item adding clear core indicators for shared resources,
\item using isolation-friendly synchronization mechanism,
\item implementing programming features that are helpful for program analysis and verification.
\end{inparaenum}
We also discuss our experience of interacting with the Linux community.
We believe that our experience sharing is valuable to application developers, OS designers, and system researchers.

With years of development and improvement, we significantly reduce cross-core performance interference in Linux. 
Using {\tt cyclictest}~\cite{cyctest}, we demonstrate a 8.7$\times$ reduction in the worst-case response time.
The reduced interference results in smaller jitter and up to 11.5$\times$ better general schedulability.
We further use the core Flight System (cFS)~\cite{cfs_web} and Robot Operating System 2 (ROS2)~\cite{ros2} to evaluate the end-to-end system performance, and we achieve 1.6$\times$ and 1.64$\times$ lower worst-case latency than RT-Linux.
Most of our modifications have been merged into Linux upstream since the 4.19 kernel and released in commercial distributions.

The contributions of this paper are:
\begin{itemize}
    \item Drawing attention to an important research problem that the operating system itself has become a primary source of timing interference from an industrial perspective.
    \item Sharing our industry practice of identifying and fixing a multitude of cross-core interference issues in Linux.
    \item  Summarizing general principles for mitigating interference of OSes and discussing lessons for future research.
    \item Providing a comprehensive evaluation of our improvements with schedulability tests and real-world use cases.
\end{itemize}


%% file: tex/motiv.tex
\begin{figure}[!t]

  \centering
  \includegraphics[trim={0 0 0 1cm},width=0.465\textwidth,clip]{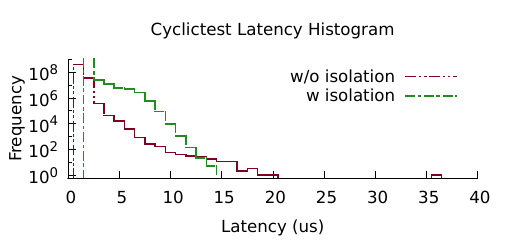}
  \vspace{-0.5em}
  \caption{\small Latency distribution of {\tt cyclictest} on an idle system.
  }
\vspace{-1.5em}
  \label{fig:moti_idle}
\end{figure}

\section{Motivation and System Model}
\label{s:motiv}

\subsection{Existing Interference Isolation Mechanisms in Linux}
\label{ss:linux}

The Linux community develops various mechanisms to improve the cross-core interference isolation.
However, these mechanisms have two major flaws.


\head{Difficult to Operate and Maintain.}
Table~\ref{tbl:iso} summarizes various performance isolation techniques in existing Linux.
These mechanisms involve a wide range of aspects, from kernel compilation parameters, boot parameters to user-level tools.
Such diverse mechanisms bring the following challenges to system operation and maintenance.
First, the maintainers need to be aware of all the mechanisms and apply them correctly to achieve the desired isolation.
Even experienced engineers may overlook some isolation measures, leading to unexpected jitter and increased worst-case execution time of products after deployment.
Second, it is difficult to locate the root causes when performance interference occurs.
The root cause may be due to that the existing mechanism not covering it, or the mechanism has bugs or is not correctly used. 


\head{Incomplete Interference Isolation.}
Due to the complexity of the kernel code and the incompleteness of various isolation approaches, the current system still faces various performance interference issues.
These problems are challenging to discover in the design stage, and can cause deadline misses and long tail latency during operation.
This seriously threatens the time safety of real-time systems or latency-sensitive applications.

\begin{table}[!t]
   \centering
   
  \caption{\small Measured maximum latency of {\tt cyclictest}.}
   \renewcommand\arraystretch{0.9}
  \begin{tabular}{cccc}
    \toprule
    System         & Linux  & RT-Linux & \cellcolor[gray]{0.8}openEuler  \\ \midrule
    Maximum Latency & 104 us & 48 us    & \cellcolor[gray]{0.8}12 us \\ \bottomrule
  \end{tabular}
  \label{tbl:cyc}
   \vspace{-1.5em}
\end{table}

\begin{table*}[t]
\scriptsize
  \centering
    \caption{\small Existing Linux isolation mechanisms that try to minimize the performance interference on isolated cores.}
  \renewcommand\arraystretch{0.7}
  \begin{tabular}{ccl}
    \toprule
    Category                       & Commands & \multicolumn{1}{c}{Description} \\ \midrule
    \multirow{1}{*}{\begin{tabular}[c]{@{}c@{}}Kernel \\Compilation \\ Options\end{tabular}} & \begin{tabular}[c]{@{}l@{}}CONFIG\_NO\_HZ = y\\ CONFIG\_NO\_HZ\_COMMON = y\\ CONFIG\_NO\_HZ\_FULL = y\end{tabular}         &  \begin{tabular}[c]{@{}l@{}}Enable the tickless mode that allows the kernel to \\ reduce the number of scheduling-clock interrupts.\end{tabular}     \\ \cmidrule{2-3}
                             & CONFIG\_CPU\_ISOLATION = y      &  Enables the CPU isolation feature in the kernel.     \\ \midrule
    \multirow{8}{*}{\begin{tabular}[c]{@{}c@{}}Kernel \\Boot \\ Parameters\end{tabular}}    & {\tt isolcpus =} core\_list & \begin{tabular}[c]{@{}l@{}}Remove specified cores from the general kernel \\ SMP balancing and scheduler algroithms.\end{tabular}     \\ \cmidrule{2-3}
    &  {\tt irqaffinity =} core\_list & Specify non-isolated cores to handle IRQs.         \\ \cmidrule{2-3}
    &  {\tt rcu\_nocbs =} core\_list  & Exclude listed cores from running RCU callbacks.   \\ \cmidrule{2-3}
    &  {\tt nohz\_full =} core\_list  & Enable the full dynticks mode on isolated cores.   \\ \cmidrule{2-3}
    &  {\tt pcie\_aspm = off}         & Disables the ASPM feature for PCIe devices.        \\ \midrule
    \multirow{3}{*}{\begin{tabular}[c]{@{}c@{}}User-level \\ IRQ tool\end{tabular}}     &  {\tt systemctl stop irqbalance.service}  & Disable {\tt irqbalance} service.           \\ \cmidrule{2-3}
    &  {\tt echo X > /proc/irq/*/smp\_affinity\_list}   & Set IRQ affinity to non-isolated cores X.           \\ \midrule
    \multirow{8}{*}{\begin{tabular}[c]{@{}c@{}}User-level \\ Memory \\ Management\end{tabular}}     &  {\tt echo 0 > /proc/sys/kernel/numa\_balancing}  & Disable NUMA balancing.      \\ \cmidrule{2-3}
    & {\tt echo never > /sys/kernel/mm/transparent\_hugepage/enabled}    & Disable Transparent Huge Pages mechanism. \\ \cmidrule{2-3}
    & {\tt echo 0 > /sys/fs/cgroup/memory/memory.move\_charge\_at\_immigrate}  & \begin{tabular}[c]{@{}l@{}}Disallow moving charges when migrating  \\ tasks between {\tt cgroups}.\end{tabular}  \\ \cmidrule{2-3}
    & \begin{tabular}[c]{@{}l@{}}{\tt echo 0 > /sys/kernel/mm/ksm/merge\_across\_nodes} \\ {\tt echo 0 > /sys/kernel/mm/ksm/run}\end{tabular}        &   Disable Kernel Same-page Merging (KSM).        \\ \midrule
    \multirow{4}{*}{\begin{tabular}[c]{@{}c@{}}User-level \\ Task tool\end{tabular}}    & {\tt taskset -pc X kswapd*}  & Move kswapd thread to non-isolated cores X.   \\ \cmidrule{2-3}
    &   {\tt echo X > /sys/kernel/pcrypt/pdecrypt/serial(parallel)\_cpumask}   & Bind pcrypt tasks to non-isolated cores X.    \\ \cmidrule{2-3}
    &   {\tt echo X > /sys/devices/virtual/workqueue/cpumask}   & Bind workqueue tasks to non-isolated cores X.    \\ \midrule
    \multirow{8}{*}{\begin{tabular}[c]{@{}c@{}}User-level \\ Other tools\end{tabular}}   &   {\tt cpupower frequency-set -g performance}  & Set CPU governor to performance mode.      \\ \cmidrule{2-3}
    &  {\tt echo 0 > /proc/sys/kernel/timer\_migration}   & Disable the timer migration mechanism.  \\ \cmidrule{2-3}
    &  {\tt echo 1 > /sys/kernel/rcu\_normal}       &  Disable preemption of RCU callback.         \\ \cmidrule{2-3}
    &  \begin{tabular}[c]{@{}l@{}}{\tt echo -1 > /proc/sys/kernel/sched\_rt\_runtime\_us} \\ {\tt echo -1 > /proc/sys/kernel/sched\_rt\_period\_us}\end{tabular}       &  Disable the CPU time limit for real-time cores.   \\ \cmidrule{2-3}
    & {\tt swapoff -a}    &  Disable the swap partition. \\ \bottomrule
  \end{tabular}

  \label{tbl:iso}
\end{table*}


To measure the effectiveness of existing mechanisms, 
we apply all the mechanisms in Table~\ref{tbl:iso} to isolate 24 cores to run the \texttt{cyclictest}.
We first measure the interference on an idle system of Vanilla Linux kernel (the unmodified
Linux 5.10 kernel) with and without isolation and report the result in Figure~\ref{fig:moti_idle}.
Without enabling any isolation mechanism, the maximum latency of \texttt{cyclictest} is more than 35 us.
Enabling the isolation mechanisms significantly reduces the interference. 
However, the maximum latency is still 15 us under isolation. 
Such latency is inadequate for meeting strong timing requirements in many real-time systems (the worst-case latency should be smaller than 10 us
\footnote{10us latency is summarized according to various business requirements. For example, industry 5G URLLC pipeline requires 1-4 ms end-to-end latency~\cite{khan2022urllc,huawei-5G,samsung-5G}.
Our system is deployed in the entire communication pipeline (gateway, base station, and core networking).
By operating each stage, the latency requirement ultimately decomposed into the OS is 10us.}).


To accurately estimate the worst-case latency that can occur in practice, we introduced a significant amount of interference workload (details in Table~\ref{tbl:interf}) only on nonisolated cores, and Table~\ref{tbl:cyc} presents the resulting maximum latency of different systems.
As the current mechanisms are unable to resolve many interference issues, the measured maximum latency in Linux and RT-Linux rapidly increases to 104 us and 48 us, respectively.
In contrast, openEuler, a Linux distribution that integrates all of our kernel modifications, is capable of achieving a worst-case latency of 12 us.

In summary, due to the lack of systematic analysis and design of the system’s isolation mechanism, many design defects and potential bugs will be exposed with extensive practical product usage.
Therefore, we summarize our years of experience and lessons and provide more practical insights and case studies to support deeper research in the future.

\subsection{Hypervisor Approach}

There is a trend within the real-time community of utilizing hypervisors, with real-time operating systems (RTOSes) or Linux running on top of them. 
The two major flaws of Linux system interference isolation mentioned above can be easily overcome through hypervisors.
But in practice, many scenarios (e.g. Robot Operating System~\cite{ros2} and edge computing) still have to use Linux-based systems due to commercial concerns directly: 
\begin{inparaenum}
    \item Virtualization usually reduces performance, especially internal communication performance. For instance, Gadepalli \etal \cite{chaos_rtas_19} showed over 200\% degradation in round-trip communication and wake-up latency. 
    \item Porting effort and maintenance costs prevent the migration of many legacy systems. 
    \item Inside the virtual machine, mitigating interference is still desired.
\end{inparaenum}
Therefore, mitigating the cross-core interference within the Linux kernel is still needed.

\subsection{System Model}
\label{ss:sysmodel}

\begin{figure}[!t]
  \centering
    \vspace{-1em}
  \includegraphics[width=0.4\textwidth]{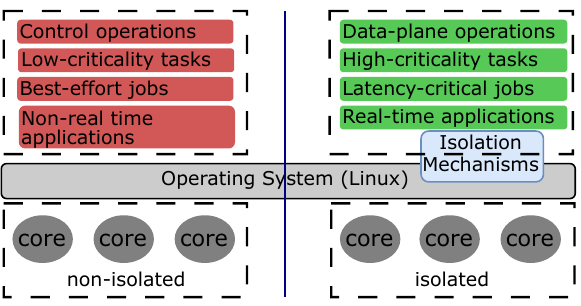}
  \caption{\small System model.}
  \label{fig:model}
  \vspace{-1.5em}
\end{figure}

We consider a multi-core processor with two partitions, $P_r$ and $P_n$.
As shown in Figure \ref{fig:model}, 
$P_r$ is reserved exclusively for real-time workload, while the rest of the cores belong to $P_n$ which performs non-real time tasks.
To ensure timing requirements, we enable all necessary isolation mechanisms to the partitioned system, including all mechanisms in Table~\ref{tbl:iso} and other available hardware partition solutions, such as memory, cache, devices, and bus bandwidth.

We make the following three assumptions on the system model.
First, applications on $P_r$ satisfy real-time constraints.
They have bounded worst-case execute time, use real-time synchronization protocol, and are scheduled by a proper scheduling algorithm.
Second, applications on $P_n$ never directly communicate and interact with tasks on $P_r$.
Between the two partitions, there is no shared memory, direct communication channel, nor other dependencies.
Third, tasks on $P_n$ do not have any real-time constraints.
These tasks can overload the CPU, perform unlimited I/O operations, and invoke arbitrary system services.
Based on these assumptions, we aim to guarantee that any interference occurring on $P_r$ is from $P_n$ and caused by the OS itself.




%% file: tex/empirical.tex
\section{Cross-Core Interference Bugs And Their Fixes}
\label{s:bug_and_fix}

We have identified and fixed 34 cross-core interference bugs in the Linux kernel in the last 6 years.
40 patches have been incorporated into the Linux mainline~\cite{arm64mm_patch,armasid_patch,perf_top_patch,perf_patch,tick_patch,nohz_patch,ticksched_patch,jiffies_patch,wkq_patch,psci_patch,irq_patch}.
One set of patches for fixing the ASID (Address Space Identifier) issue (\S\ref{ss:res_asid}) is under discussion with the community~\cite{asid_pr}.
9 issues are worked around by restricting applications behavior and system usage.
3 fixes are maintained privately due to business needs.
All fixes are deployed in our production environment and delivered to millions of servers and devices. 
We summarize cross-core interference bugs into three categories based on their functionality: task scheduling and placement, resource management and sharing, and concurrency and coordination.
They are scattered across multiple subsystems of the kernel, and the total amount of code we analyzed and investigated exceeds 80,000 lines.
The following sections discuss typical examples of different types of bugs and the reasons for the bugs.

\input{tex/task-new}

\input{tex/resources-new}

\input{tex/concurrency-new}

%% file: tex/task-new.tex
\subsection{Task Scheduling and Placement}
\label{s:design}

We notice that the scheduling of system tasks in current Linux is prone to cross-core interference bugs.
In this section, we use two typical bugs to explain how system task scheduling bugs could introduce cross-core inferences.

 \begin{figure}[!t]
   \centering
   \includegraphics[width=3.3in]{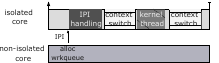}
   \caption{\small Cross-core interference of {\tt workqueue}.}
   \label{fig:kwq_i}
  \vspace{-1.5em}
 \end{figure}

\subsubsection{Indistinguishable Core Activation}
\label{ss:workq}

This interference happens when {\tt workqueue} operations mistakenly wake up worker threads on all cores, including the isolated ones.
The {\tt workqueue} is a lightweight mechanism to schedule simple kernel tasks.
Each {\tt workqueue} is a queue of work items, which is a function that is executed asynchronously by multiple worker threads.
The worker threads are usually organized into a per-core thread pool and are activated by the {\tt workqueue}'s own task scheduling policies.
The policies determine when and where to execute the work items.

\head{Redundant Activation.}
Ideally, before {\tt workqueue} sends any work item to a core, it should check if the work item is appropriate for the target core.
However, we find that {\tt workqueue} fails to make this check and thus introduces cross-core inferences.
Specifically, we identified two problems in the {\tt workqueue} implementation.
First, some operations (\eg\ \texttt{pwq\_adjust\_max\_active}) indiscriminately wake up worker threads on all cores, including the isolated ones.
Second, some irrelevant operations, such as \texttt{alloc\_workqueue}, usually activate unnecessary worker threads and thus activate isolated cores.
For both cases, an IPI (Inter-Processor Interrupt) is sent to the isolated core to wake up a worker thread.

As shown in Figure~\ref{fig:kwq_i}, core activation causes significant interference to isolated cores.
An IPI invokes the isolated core to perform four operations:
\begin{inparaenum}[1)]
\item handle the IPI, 
\item switch context to the worker thread,
\item execute the work item, 
\item switch context back to the real-time task.
\end{inparaenum}
All the operations consume CPU time and introduce interference.
IPI handling and context switching disrupt the real-time task and lead to a significant jitter.
Moreover, the worker thread execution itself introduces additional overhead to the
isolated core.


\head{Solution: Core Activation Restriction.}
To address the above problems, we propose three technical solutions to restrict the activation of worker threads on isolated cores~\cite{wkq_patch}.
First, we only wake up a worker thread on a core when it has pending work items. 
Second, we restrict the activation mechanism so that a worker thread can only wake a core within the same partition (\S\ref{ss:sysmodel}). 
It means that a non-isolated core can only send an IPI to another non-isolated core but can not send an IPI to an isolated core.
Third, we check and guarantee that irrelevant operations (e.g. {\tt alloc\_workqueue}) do not
contain any thread activation and task distribution operation.


\subsubsection{Inconsistent Task Migration}
\label{ss:interrupt}


Task migration, which means moving a process or thread from one CPU core to another, is a common operation that can be triggered by either users or the OS.
The central design of a migration policy is to determine which task to migrate and the migration destination.
Migrating a task involves various components from the kernel to user-level tools, and any mismatch between the migration components can lead to inconsistency and interference.

\head{Isolation-Agnostic Migration.} 
A typical interference case of inconsistent migration is as follows.
A task is initially placed on isolated cores due to incorrect core selection.
As the task executes, it spawns other tasks on the same cores.
When the OS or a user detects the misplacement and tries to migrate the task, the migration interface may not apply to the spawned tasks.
These spawned tasks are left on the isolated cores, leading to interference.
For instance, some microphone driver pins its IRQ threads by using  \texttt{cpumask\_local\_spread} function.
This function returns a core but does not check the isolation status.
When the IRQ thread is later moved to a non-isolated core, the spawned kernel threads remain on the isolated core.
Moreover, the user-level utility function {\tt irqbalance} also ignores the isolation configuration of cores when there is no explicit user request.


\head{Solution: Isolation-Aware Selection.}
To include the isolation status in the task migration, we implement two fixes to select the target core for migration.
First, for the \texttt{cpumask\_local\_spread} function, we modify the core selection logic to
check the isolation status.
We only select a core from the online non-isolated cores.
Only if all non-isolated cores are offline will we return an isolated core.
Second, for the \texttt{irqbalance} function, we check if the user explicitly provides a core list for balancing~\cite{irqbalance_patch}.
If not, we follow the OS default isolation configuration that excludes the isolated cores specified in the boot parameters.

%% file: tex/resources-new.tex
\subsection{Resource Management and Sharing}
\label{s:resource}


We find that another source for the cross-core interference bugs is the incorrect management of shared resources across different cores.
Specifically, in this section, we discuss two typical bugs related to resource management.

\subsubsection{Hardware Resource-ASID}
\label{ss:res_asid}


Address Space Identifier (ASID) is a technique to reduce the overhead of TLB flushes and context switches. 
An ASID is a unique identifier assigned to each process and is used to differentiate between virtual address spaces.
The OS uses ASID to identify the correct mapping and does not flush all the TLB entries when switching contexts. 
The number of ASIDs supported by the processor is limited (\eg\ ARM64 supports 65536 ASIDs~\cite{armdeveloper}), and all ASIDs are shared by all the cores.
When the number of allocated ASIDs exceeds the limit, the kernel needs to handle ASID reuse between processes.
However, the existing ASID design has inappropriate ASID allocation that causes subtle cross-core interference during ASID exhaustion and reuse.

Existing ASID management follows a lazy manner. 
A global bitmap is used to record the availability of each ASID.
When a process is created, the system allocates a fresh ASID, but it \textit{does not} release the ASID when the process exits.
After all the ASIDs are exhausted, the system reclaims and reallocates the ASID. 
To reuse and reallocate ASIDs, the system records the current version number of ASID by maintaining a global variable, \texttt{asid\_gen}.
Each process maintains a local variable, \texttt{local\_gen}, to record the \texttt{asid\_gen} at the time the process was assigned an ASID.
Upon each context switch, the process compares \texttt{asid\_gen} and \texttt{local\_gen} to judge whether the current ASID of the process has been reallocated. 
Since the same ASID may have been allocated to another process, the system needs to force TLB flush.

\begin{figure}[!t]
  \centering
  \includegraphics[width=3in]{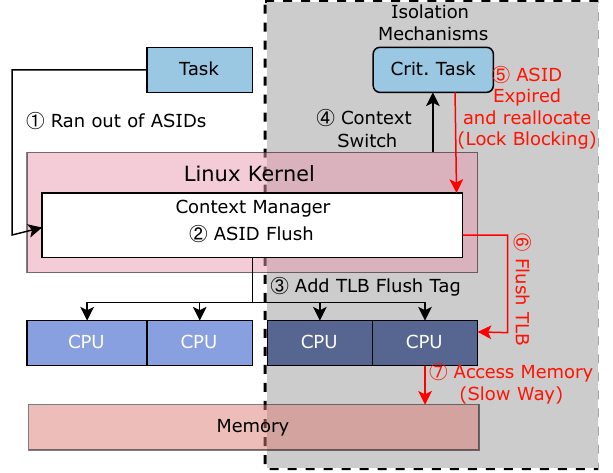}
  \caption{\small The process of inter-core interference caused by ASID exhaustion and TLB refresh.}
  \label{fig:asid_intr}
 \vspace{-1.5em}
\end{figure}

\head{Shared ASID Exhaustion.}
The existing implementation does not consider that the ASID is a shared resource among all the cores.
The exhaustion of ASIDs in non-isolated cores is also for isolated cores, which will lead to ASID reallocation and TLB flush.
Figure~\ref{fig:asid_intr} shows the interference path of ASID exhaustion.
When the non-isolated core exhausts all ASIDs (\ding{172}), the system starts to flush and reallocate ASIDs (\ding{173}) and marks the TLB flush tag of all cores (\ding{174}).
When a context switch occurs in the isolated core (\ding{175}), the process finds its ASID expires and requests a new one.
Because the ASID operations (\eg\ allocation and flush) require modifying the global bitmap, a spinlock is used for synchronization and blocks the operation on the isolated core, which is an undesired interference (\ding{176}).
If a new ASID is allocated, the isolated core has to flush the TLB (\ding{177}).
Moreover, after the process resumes, it suffers TLB miss when accessing memory (\ding{178}).
It significantly slows down the process execution.

  


\head{Solution: ASID Isolation.}
We propose to partition the ASID space and integrate ASID into the core isolation~\cite{asid_pr}.
Given the predictability of real-time applications, both the process count and creation rate are bounded.
Based on this observation, we reserve enough ASIDs to the isolated cores and guarantee non-isolated cores never use reserved ASIDs.
By separating the ASID allocation and management of real-time processes from non-isolated cores, we can avoid ASID interference. 
The similar idea also applies to x86 which has the same mechanism but fewer IDs than ARM.

While the ASID space is shared across all cores, we divide the ASID space into two subspaces. Isolated cores and non-isolated cores use different subspaces.
Each subspace manages an independent ASID range, thus the lock and generation counter are also managed separately. 
In ASID management operations, We first check whether the current core is isolated.
Then, all subsequent operations are restricted to the corresponding subspace.
The basic process of each ASID operation is similar to the existing implementation, except that we replace global variable access with subspace-wide variables.
With our ASID isolation approach, we can guarantee that non-isolated cores never interfere with isolated cores.
For example, if non-isolated cores exhaust ASID in their subspace, they will reallocate ASID exclusively from its assigned subspace.
ASIDs assigned to the isolated cores remain unaffected, and non-isolated cores will neither compete for the lock nor mark the TLB flush flag on isolated cores.

\subsubsection{Device Management}
\label{ss:device}

The OS needs to clear the stale status of a device when it is updated during plug-in, probe, or removal phases.
In a general system, a device is often shared by all the cores.
Updating the device status will invoke other cores to clean up the core-local device status. 
However, such cross-core processing is undesirable in real-time systems since it triggers cross-core interferences.
Particularly, real-time applications on isolated cores explicitly avoid sharing devices with
non-isolated cores, thus no status need to be maintained on isolated cores at all.

\head{Unnecessary Backlog Flushing.}
We take network interface controller (NIC) management as an example.
A NIC has a per-core {\tt backlog} queue that holds pending packets.
When uninstalling a NIC, the system flushes all backlog queues to prevent missing packets, which cause interactions over all the cores.
However, the current implementation does not consider whether the flush is necessary or not, and does not distinguish between isolated and non-isolated cores.
This causes obvious cross-core interference even though the NIC is never used by the isolated cores.

\head{Solution: On-demand Flush.}
The solution to prevent interaction from device management is to introduce a flush operation that is triggered only when necessary.
Before flushing a remote core, we check if the target core has any pending packets.
If the target core has no pending packets, we skip the flush operation.
Otherwise, we check the isolation status of the caller core.
If the caller core is isolated, we complete the flush operation since the flush is necessary and is not external interference.
If the caller core is not isolated, we report a warning and record the problematic behavior.






%% file: tex/concurrency-new.tex
\subsection{Concurrency and Coordination}
\label{s:concurrency}


On multicore processors, concurrency and synchronization are inevitable. 
Especially when the single OS kernel is shared by isolated and nonisolated cores, synchronization between them is required.
For example, different cores have to synchronize their clocks.
Linux currently employs various synchronization mechanisms that introduce different degrees and types of cross-core interference.
We classify and examine these mechanisms and show that each category has different interference problems.


\subsubsection{Lock-based Synchronization}
\label{ss:jiffes}

Jiffies is a globally shared variable to record the number of timer ticks lapsed in the system.
Therefore, accessing and updating the jiffies requires synchronization across multiple cores.
The current implementation employs a combination of multiple methods, including disabling
interrupts and using read-write lock ({\tt seqlock}).

\head{Concurrent Blocking.}
The jiffies synchronization inevitably causes interference on isolated cores.
When reading the jiffies value, the isolated core may encounter an unbounded retry loop due to conflicting modifications from non-isolated cores.
If the isolated core is a writer, it may also be blocked by concurrent writers on
non-isolated cores.
Moreover, the spin lock does not follow any real-time locking protocol~\cite{spin_tc_23}.

\head{Solution: Compress Critical Sections.}
Our fix mitigates the interference by reducing the critical section length~\cite{jiffies_patch,ticksched_patch}, which decreases the execution time of each individual interference and lowers the probability of being interfered with.
Inside the {\tt tick\_do\_update\_jiffies64} function, we remove unnecessary changes of the
atomic counter when the jiffies is up to date.
We also identify a part that does not need {\tt seqlock} protection, thus reducing the time of holding a lock.

\subsubsection{Distributed Coordination}
\label{ss:vmstate}

We present a case of message-based coordination for statistics maintenance in the system.
The system has a large number of statistics related to various aspects of virtual memory, such as memory usage and disk scheduling.
Each core maintains its local statistics, and the system aggregates data from all the cores on demand and checks for any possible inconsistencies.
For example, the {\tt vmstat\_shepherd} function schedules work on all cores to update the virtual memory statistics.
To avoid uncontrolled interference from statistics updating, we provide additional interfaces that only process statistics within a given partition and ensure that the caller is also a member of that partition.






%% file: tex/lessons.tex
\section{Lessons}
\label{s:lessons}
Regardless of the countless man-months spent fixing cross-core interference bugs in Linux, we despairingly found that it is impossible to fix all cross-core interference bugs in Linux.
The key reason is that Linux does not have a well-designed mechanism to enforce isolation between cores.
In this section, we summarize the pitfalls in the Linux design that make it difficult to fix cross-core interference bugs.
We also provide suggestions on how to design an interference-free operating system for future systems.
In the last, we discuss our experience of interacting with the Linux community.

\subsection{Lesson 1: Unified Isolation Mechanisms and Interfaces}
\label{ss:les1}

In Linux, isolation between cores is ensured ad hoc: each task subsystem, resource manager, and synchronizer manage isolation between cores independently; and they are closely entangled together, making the whole system prone to bugs and inconsistencies.
For example, Linux has multiple independent core isolation mechanisms  (\eg\ {\tt isolcpus}, {\tt irqaffinity}, {\tt task affinity}, {\tt cpu\_mask} and {\tt nohz\_full}) that are orthogonal and have distinct methods for managing the isolation status of cores, making it even more complex to correctly avoid cross-core interference.
The only way to avoid interference from non-isolated cores to isolated cores is to check the isolation status for every isolation mechanism, which is error prone.
Even worse, different maintainers in the community have different understandings about these mechanisms.
For example, the community fixes a bug similar to the cases discussed in \S\ref{s:design}: starting a kernel thread does not consider the core isolation indicated by the {\tt isolcpus} boot parameter~\cite{linux_patch_isolcpu}.
However, this fix ignores {\tt nohz\_full} which is a strong indicator that the core may execute latency-critical tasks, but the community cannot reach an agreement on whether {\tt nohz\_full} related to isolated CPUs or not.

Worse still, it is difficult to detect and fix cross-core interference bugs due to
 the code and engineering complexity in existing Linux.
We performed an empirical study. 
In Linux 5.10.210, there are at least 19,668 lines of code
\footnote{
This number is a significantly low estimate and only includes the amount of code for functions that directly run and control these kernel threads.
This is the same as our statistic result in resource management.
}
for kernel thread scheduling that manages at least 306 kernel thread tasks.
More than 20 {\tt workqueue}s and 140 {\tt cpu\_mask}s are used and are scattered across different modules covering more than 100 files.
Any inconsistency between these isolation mechanisms will cause cross-core interference.
For resource management, there are at least 10659 types of drivers and various hardware resources, including the design of memory, cache, bus, and other aspects such as ASID, cache tags, and page table walk cache, many of which are not easily discovered by us.
Even worse, although resources or tasks are managed by dedicated management components, the code for the management components is tightly entangled.
We find many of the management components have dependencies on another component.
For instance, there are at least 221 kernel modules and drivers that use {\tt workqueue} to manage tasks.
This significantly increases the complexity of debugging.


\head{Suggestions.}
We argue that it is necessary to design a unified mechanism and interfaces to reduce bugs.
First, we need a unified way to check and manipulate the isolation status, instead of using multiple independent mechanisms and checking the isolation status of each mechanism individually.
Second, we need to have a unified interface for operations that may use cores.  
For task scheduling, we need a unified interface that checks the status of the target core and the necessity of invoking isolated cores.
For resource management, the unified interface needs to prevent non-isolated cores from operating the resources belonging to or shared by the isolated cores.

\subsection{Lesson 2: Clear Indicators for Resources on Isolated and Non-Isolated Cores}
\label{ss:les2}

Another challenge that prevents developers from avoiding cross-core interference is that the Linux kernel lacks a clear boundary between resources of isolated and non-isolated cores. 
Thus, it is difficult to judge whether using a resource may cause inter-core interference.
For example, the main reason for the two bugs in \S\ref{s:resource} is that the ASIDs and \texttt{backlog} buffers do not have any indicators for shared usage from isolated cores.
Therefore, the developers may overlook the isolation status and introduce inter-core interference. 

Note that there are many similar shared resources between cores causing cross-core interference.
For example, shared tags for the last-level cache (not the cache itself)~\cite{hisi_tags} may also cause cross-core interference bugs.
Another example is interrupt handling.
We notice that interrupts generated by PCI devices are dispatched to isolated and non-isolated cores indiscriminately. 
It is particularly challenging to detect and fix all these bugs in practice.
For example, each I/O device has a per-core buffer, which may also suffer problems similar to the \texttt{backlog} bug in \S\ref{s:resource}.
Allowing each driver or resource management component to individually manage core isolation status is error prone and may introduce conflicts.

\head{Suggestions.}
We suggest that all system resources, events, and interrupts should have a clear core indicator about which core can they be used on. 
These information needs to be fully exposed to developers.
In this way, they can use these information to avoid implement functions that may lead to cross-core interference (for example, there is competition for shared resources between isolated and non-isolated cores).

\subsection{Lesson 3: Isolation-Friendly Synchronization Mechanism}
\label{ss:les3}

In existing Linux, it is almost impossible to completely eliminate synchronization between cores. 
Thus, we can hardly avoid the interference introduced by synchronization.
Our current effort alleviates the interference, but it cannot fully address the interference because we did not choose to completely redesign the synchronization system. 
We choose to stay with the current synchronization mechanism because many system logic are deeply coupled with synchronization mechanisms. 
Modifying synchronization mechanisms can probably cause bugs and break normal functional logic. 
For example, we chose to conservatively fix jiffies instead of changing the synchronization mechanism or deprecating it. 
This is because that although current systems are now equipped with high-resolution timers, many old kernel codes and old device drivers still rely on the jiffies for clocking.

\head{Suggestion.}
To completely avoid interference caused by synchronization, we suggest adopting advanced kernel design patterns, such as the quiescence-based microkernel~\cite{speck_rtas_15}. 
Its lock-free design offers controlled bounds on kernel execution and synchronization time, avoiding concurrency interference caused by kernel-resource synchronization.

\subsection{Lesson 4: Verifiable Programming Practices}
\label{ss:les4}

We have suggested using unified isolation mechanisms and interfaces, and clear isolation indicators for resources.
However, we also notice that developers may still make mistakes due to the complexity of the Linux Kernel.
To this end, we suggest designing the isolation interfaces, mechanisms, and indicators in a way that is friendly for formal verification and automated program analysis. 

First, we suggest adding an isolation identifier to the function names in the kernel.
This will allow the taint-based program analysis techniques~\cite{kennedy1979survey,flow_pldi_14} to identify interference for non-isolated cores to isolated cores.
In practice, we can set the data from the non-isolated cores to be tainted and set the operations to isolated cores as sensitive locations.
Then, we can use a standard taint analysis technique~\cite{drtaint_sec_17,taintpipe_sec_15} to detect if the data from non-isolated cores can reach operations on isolated cores. 
By having a clear identifier on the functions, the taint analysis can automatically identify all data from non-isolated cores. 
Second, we suggest increasing more fine-grained trace operations. 
For example, the trace of accessing shared resources. 
This practice will allow us to have more control over the behavior of systems. 
For instance, we can utilize trace technology and formalized analysis methods in combination \cite{de2020demystifying}. 
We need to trace all system events that may have an impact on latency, including their running positions and the maximum latency observed. 
These attributes are utilized in the analysis to determine the sources of interference and the boundaries of latency time within the Linux kernel. 

Linux, as a general-purpose monolithic kernel, is almost impossible to realize  full formal verification.
The above suggestions are merely ways to help with the verification of some functionalities.
To enhance the verifiability, it is necessary to draw on the design philosophy of formally verified OS (\eg\ seL4~\cite{klein2009sel4} and CertiKOS~\cite{certikos_osdi_16}) to redesign the kernel. 
This specifically includes the following aspects:
\begin{inparaenum}[1)]
\item Further modularize and separate the components in the kernel space to reduce verification complexity.
\item Adopt stricter abstraction and encapsulation in terms of kernel interfaces and system calls to reduce dependencies between modules.
\item Provide more fine-grained access control permissions.
\item Introduce formal specifications for key components, such as the noninterference specifications~\cite{nelson2020noninterference} that ensure the resources of high-priority users are not affected by low-priority users.
\end{inparaenum}

\subsection{Linux Community Interaction}
\label{ss:community}

Collaboration with the Linux community is an important part of our work.
We share some experience and issues in collaborating with the community that maybe helpful to the community evolution and researcher engagement.

\head{Insufficient participation from academia.}
The Linux community is mainly dominated by engineers from large companies.
For example, more than half of the patches in Linux 6.1 come from five major companies~\cite{dev_lwnet}.
Insufficient academic involvement poses challenges to the development of Linux-based real-time systems, as they require theoretical foundations and research analysis from academia.
Without academic support and theoretical verification, design defects inevitably cause subtitle cross-core interference bugs (\S\ref{s:bug_and_fix}) which are difficult to detect and fix.
Therefore, it is necessary to encourage more scholars to actively participate in the community to discuss and review the system design.
Additionally, the community needs to be more open and establish effective collaboration mechanisms to support academic participation.
For instance, the Linux community should fund more academic collaboration projects and set up a dedicated academic committee to review and discuss real-time-related designs.
The openEuler community is doing some similar practices.

\head{Production driven development.}
Most contribution to the Linux community is predominantly focused on the specific product that the developers works on, thus the completeness and generality of system design are often neglected.
This is particularly vulnerable to real-time systems as an unanticipated interference bug can significantly impact the system timing safety.
The typical process of companies contributing to the community involves
\begin{inparaenum}[1)]
\item identifying a problem in the product, 
\item fixing the issue tailored to the product, 
\item testing and verifying the fix, 
\item deploying to the product, and 
\item finally pushing the modifications to the community.
\end{inparaenum}
This process is reasonable for a profit-oriented company but brings some adverse effects on the community.
\begin{inparaenum}[1)]
\item The fixes may be incomplete or introduce new issues, and the committers cannot always accurately review fixes to find potential issues due to limited production experience.
For example, a previous core isolation fix was reverted~\cite{linux_patch_revert} and we have to fix it privately instead of using the community solution.
\item Some merged fixes may impact the maintenance of other companies. 
For some interference bugs we detected, the community merges other fixes first, so we have to modify our solution and adapt to the community changes to ensure compatibility and reduce private patches.
\item Engineers lack incentives to improve their patches.
Since these fixes have already addressed the product issues and been deployed in products, it is neither worthy nor safe to spend extra time improving the solution.
For instance, based on multiple iterations of discussions and feedback from the community, our ASID fix (\S\ref{ss:res_asid}) is desired to support dynamic partitioning, but the significant effort required brings more work pressure to engineers.
\end{inparaenum}
Completely solving these issues is hard, but we hope that involving more academic researchers can effectively mitigate these issues.

\subsection{Development and Operational Experience}

While we provide some high-level advice on reducing interference from \S\ref{ss:les1} to \S\ref{ss:les4}, we also have the same problems as \S\ref{ss:community}.
Our actual fixes in the kernel are still fragmented and developed in an ad-hoc way, as we must avoid intrusive modifications to deliver stable productions.
However, our experience, principles, and analysis reveal general problems for cross-core interference and can guide and be applied to other systems.
For example, we are developing a new micro-kernel-based commercial OS, and the experience we have accumulated in Linux helps us avoid many interference issues in the new OS.
Additionally, as we discussed in \S\ref{ss:linux}, correctly configuring and operating the system is critical but difficult to ensure isolation.
Table~\ref{tbl:iso} merely lists significant means, not an exhaustive list.
Consequently, we suggest creating an open-source cheatsheet for the industry, which aims to record and share essential system configurations and operations across diverse real-time scenarios.


%% file: tex/eval.tex
\begin{table}[!t]
   \centering
 \caption{\small Interference workload running on non-isolated cores.}

  \begin{tabular}{cl}
    \toprule
    Name         & \multicolumn{1}{c}{Description}  \\ \midrule
    k\_workqueue & Repeatedly allocate kernel workqueues.  \\ 
    u\_thread    & Create a large number of user-level threads.  \\ 
    u\_fork      & Repeatedly fork an empty process to consume ASID.  \\ 
    nic          & Repeatedly install and uninstall an unused NIC.  \\ 
    timer        & Create a large amount of timers.  \\ 
    perf\_stat   & Run above task under {\tt perf} monitor.  \\ 
    calc\_pi &   Calculate $\pi$ in {\tt stress-ng} testsuite.  \\ 
    iomix        & Disk stressing test in {\tt stress-ng}.  \\ 
    fault        & Generate page faults in {\tt stress-ng}. \\ 
    vm           & {\tt mmap} and {\tt munmap} test in {\tt stress-ng}.  \\ 
    sock         & Socket stressing test in {\tt stress-ng}.  \\ \bottomrule
  \end{tabular}
  \label{tbl:interf}
\end{table}

\section{Evaluation}
\label{s:eval}

Our evaluation aims to answer three research questions:
\noindent\emph{Interference Mitigation (\S\ref{ss:mirco}).}
We want to study the effectiveness of our techniques to mitigate the worst-case cross-core interference.
    
\noindent\emph{Schedulability Benefit (\S\ref{ss:sim}).}
We want to assess the schedulability benefits of the system after the interference is mitigated. 
    
\noindent\emph{End-to-End Response Time (\S\ref{sec:endtoend}).}
We want to study the end-to-end response time for real-time systems in the production environment and see the improvement of our techniques. 



\head{Experimental Setup.}
We conduct all our experiments except for schedulability simulation on an ARM Kunpeng-920 server with four NUMA nodes.
Each NUMA node has 24 cores @ 2.6GHz, resulting in 96 cores in total.
We isolate the last 24 cores for real-time applications, and use the other 72 cores for non-real time tasks and interference workload.
We compare three different systems:
\begin{inparaenum}[1)]
\item Vanilla Linux – the unmodified Linux 5.10 kernel.
\footnote{
As most of our work has already been merged into the Linux mainline (\S\ref{s:bug_and_fix}), it is meaningless to compare with the latest Linux kernel that includes our improvements.
We use Linux 5.10 as the baseline because it is the last stable version before our work was massively merged.
}
\item RT-Linux – Vanilla Linux applied with the real-time Linux patches ({\tt PREEMPT\_RT} of version patch-5.10.201-rt98).
\item openEuler – a Linux distribution that integrates both RT-Linux and all our kernel modifications discussed in \S\ref{s:bug_and_fix}.
\end{inparaenum}
We apply all the isolation mechanisms listed in Table~\ref{tbl:iso} to all the above studied systems. 
We have also partitioned resources as much as possible and only focused on the OS as the interference source. 
For instance, we partitioned memory and cache by binding different NUMA nodes and avoiding memory sharing and contention.


\begin{figure}[!t]
  \centering
  \includegraphics{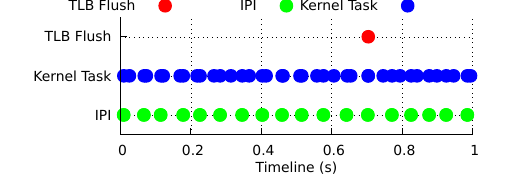}
  \vspace{-1em}
  \caption{\small interference captured on an idle isolated core.
  }
    \label{fig:moti_iso}
 \vspace{-1.5em}
\end{figure}

\subsection{Interference Workloads}
\label{ss:interfere}
To stress the kernel and obtain the worst-case performance of real-time applications, we introduce a fair amount of interference workloads on the non-isolated cores.
Table~\ref{tbl:interf} describes all the interference workloads we used in our experiments.
The workload includes both interference benchmarks and a standard stressing benchmark {\tt stress-ng}~\cite{stressng}.
These workloads have different types and stress different parts of the system, such as task scheduling, timer management, disk I/O, CPU computation, and kernel execution.
We manually bind these workloads to different non-isolated cores, so we can generate maximum interference.
Furthermore, these workloads never directly interact or communicate with applications on isolated cores. The kernel is the only shared medium between them.
As a result, our interference workloads satisfy the assumptions discussed in \S\ref{ss:linux}, and are able to reveal the isolation issues in the kernel.

To understand the amount of interference our workloads generate, we use {\tt ftrace} to record system events on an isolated idle core.
From the trace, we extract different types of interference, such as IPI handling, kernel task execution (including context switch), and TLB flush.
Figure~\ref{fig:moti_iso} visualizes the timeline of interference from 1-second trace.
As shown in the figure, IPI and kernel tasks are the major sources of interference.
TLB flush is only captured once.
This is because
\begin{inparaenum}[1)]
\item each core just needs one TLB flush after ASID exhaustion, and 
\item TLB flush is performed upon a context switch, which is rare on an idle core.
\end{inparaenum}
For a core running real-time applications, the TLB flush is much more severe.
In the following sections, we study how these interference affect applications running on the isolated cores.

\subsection{Worst-Case Interference}
\label{ss:mirco}

This section evaluates the worst-case interference and its impact on task activation on a set of micro-benchmarks.
We use two widely used tools for real-time system, {\tt cyclictest}~\cite{cyctest,cyctest_code} and {\tt oslat}~\cite{oslat, oslat_code}, to conduct the experiments.
\texttt{cyclictest} periodically blocks and wakes up a thread, and calculates the difference between the scheduled and actual wake-up time.
This time difference represents the jitter of a task activation.
\texttt{oslat} runs a busy loop with workloads and calculates the thread latency.
Both tools repeatedly track and measure the jitter, and report various statistic values, including the minimum, maximum, and average latency.

\begin{figure*}[!t]

 \begin{minipage}[t]{0.5\linewidth}
    \centering
    \includegraphics[width=\linewidth]{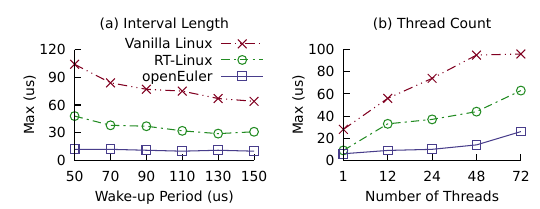}
  \caption{\small Measured maximum latency of {\tt cyclictest}.}
  \label{fig:cyc_re}
 \end{minipage}%
  \begin{minipage}[t]{0.5\linewidth}
    \centering
    \includegraphics[width=\linewidth]{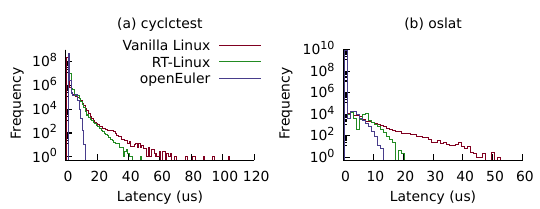}
  \caption{\small Latency distribution of {\tt cyclictest} and {\tt oslat}.}
  \label{fig:cyc_oslat_distr}
 \end{minipage}%

     \vspace{-1em}
\end{figure*}



\head{Results.}
Figure~\ref{fig:cyc_re} plots the measured latency aggregated from all testing threads.
In the figure, we adjust the wake-up interval and the number of testing threads to study the maximum activation latency under different environments in {\tt cyclictest}.
In Figure~\ref{fig:cyc_re} (a), we fix the number of threads of each isolated core to one and adjust the wake-up period from 50 us to 150 us.
We can observe that the worst-case latency of all systems increases with the wake-up period. 
This is because a longer period leads to a smaller probability that an interference happens to occur in the thread activation phase.
Thus, this does not mean the actual worst-case latency gets better as the wake-up period increases.
As a result, the worst-case latency of 50 us reflects the ground truth (listed in Table~\ref{tbl:cyc}).
For 50 us, our solution (openEuler) achieves the lowest latency.
OpenEuler reduces the worst-case latency by a factor of 8.7 and 4.0 over vanilla Linux and RT-Linux, respectively.
Figure~\ref{fig:cyc_re} (b) reports the worst-case latency with an increasing number of threads under 50 us wake-up period.
When there are more than 24 threads, some threads have to be placed on the same core.
This causes extra scheduling delay, not cross-core interference.
Hence, the measured worst-case latency goes up after 24 threads.
We can observe that, with smaller thread count, both vanilla Linux and RT-Linux have increasing latencies.
This is still because more threads have a higher probability to capture the interference. 
In contrast, openEuler is able to maintain a similar worst-case latency and is 6.78 and 3.14 times lower than vanilla Linux and RT-Linux when there are 48 threads, respectively.



In Figure~\ref{fig:cyc_oslat_distr} we show the latency distribution of {\tt cyclictest} and {\tt oslat}.
For {\tt cyclictest}, we set the wake-up period to 50 us and the number of threads to 24.
For {\tt oslat}, we use the default setting (24 cores, max RT priority, no workload, with preheat).
Because we can avoid most interference of the kernel, we can observe that openEuler not only achieves the minimum worst-case latency, but also has fewer frequencies at the tail.
For instance, we reduce the frequency of {\tt cyclictest} above 5 us latency by 25\% and 34\% over vanilla Linux and RT-Linux, respectively.
For {\tt oslat} above 5 us latency, we reduce the frequency by 61\% and 72\% over vanilla Linux and RT-Linux, respectively.


\subsection{Schedulability Tests}
\label{ss:sim}


Schedulability improvement is one of the most important benefits of interference mitigation. 
Interference can affect the schedulability of nearly all scheduling algorithms, analysis methods, and application patterns.
In this part, we evaluate the schedulability in two aspects: the \textbf{scheduling
algorithms} and the \textbf{locking protocols}.
We use the SchedCAT framework~\cite{schedcat} under different system configurations to analyze the
schedulability of the system.
We use the task activation jitter as a conservative model of the cross-core interference.

\begin{figure*}[!t]
 \centering
    \includegraphics[width=1\textwidth]{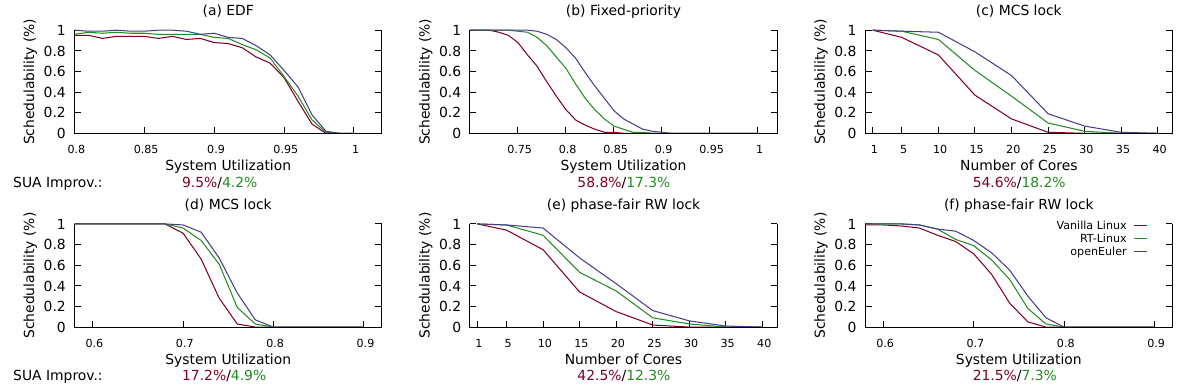}
  \caption{\small System schedulability of different scheduling algorithms and lock protocols.
  }
  \label{fig:sched_re}
\vspace{-1.5em}
\end{figure*}

\head{Task Set Generation.}
For each configuration, we randomly generate 500 task sets.
Each task set is generated by the generator proposed by Emberson et al.~\cite{emberson10techniques}.
The execution costs are calculated based on system utilization and task periods.
Periods are uniformly selected from $[10, 100]$ ms.
Each task has an implicit deadline equal to its period.
All tasks have a release jitter, which is computed from the maximum scheduling delay observed in the {\tt cyclictest} benchmark in \S\ref{ss:mirco}.
Table~\ref{tbl:cyc} shows the concrete values of release jitter for each system.
The values are measured in the setting of one thread per core with 50 us wake-up period.


\head{Scheduling Algorithms.}
We first investigate the system schedulability with two scheduling algorithms: EDF (Earliest-Deadline-First) and fixed-priority scheduling.
We use a platform of 20 cores and set the number of tasks to 40.
All tasks are independent of each other without any shared resources among them.
We use the classic RTA Analysis from Audsley et al.~\cite{audsley93newsched_theory} to analyze the
schedulability.
EDF scheduling greedily selects the task with the earliest deadline to execute.
Fixed-priority scheduling statically assigns a priority to each task.
The smaller the task's period is, the higher the priority is.

Figure~\ref{fig:sched_re} (a) and (b) shows the results of EDF and fixed-priority scheduling, respectively.
For EDF, when the task set utilization is less than 90\%, all systems are schedulable. 
However, higher utilization causes some tasks to miss deadlines.
While RT-Linux performs better than vanilla Linux due to the {\tt PREEMPT\_RT} patches, it still has many interference issues.
For openEuler, it has the smallest jitter (cross-core interference) and achieves the highest schedulability in all configurations.
For instance, with 95\% utilization, openEuler has 15\% and 7\% higher schedulability than vanilla Linux and RT-Linux, respectively.
For fixed-priority scheduling, openEuler has similar superiority and has the highest schedulability in all configurations.
It realizes up to 11.5$\times$ and 3.5$\times$ improvement over vanilla Linux and RT-Linux, respectively.



\head{Locking Protocols.}
In this scenario, tasks need to employ real-time locking protocols for synchronization.
The blocking time is a major factor to impacts the system schedulability.
We model the cross-core interference as task release jitter.
The interference occurs when a task holds a lock and other tasks wait for this lock.
This interference significantly increases the blocking time of other tasks.
We study two types of locks: the MCS lock~\cite{mcs} and the reader-writer lock~\cite{brandenburg2010spin}.
For MCS, each task has a unique lock.
For the reader-writer lock, each reader has a unique lock, and all writers share a single lock. 

Our experiments use $ \{1, 5, 10 \dots 40\}$ cores, and vary the per-core utilization across $[0.6, 0.9]$ in a step of $0.05$.
We set the default number of cores to $20$, and the default per-core utilization to $0.75$.
We generate task sets with $10$ tasks per core and use fixed-priority scheduling as the scheduling algorithm.
Only one global resource is shared among all tasks.
Tasks can either access (reader) or modify (writer) the shared resource.
Readers and writers are uniformly assigned to tasks, and they issue one read/write request per period.
The duration of read/write requests is set to $100 us$.
We use ILP-based approach~\cite{brandenburg13spin} to compute the blocking time.

Figure~\ref{fig:sched_re} (c) and (d) show the schedulability of different core counts and task set utilization of MCS lock.
Since MCS lock does not distinguish between readers and writers, all tasks will be blocked when accessing the shared resource, the schedulability of all systems rapidly decrease due to the more series contention.
Despite of the large blocking time, the cross-core interference still considerably impacts the schedulability.
We can observe that openEuler achieves the highest schedulability in all cases. Notably, with 20 cores and 75\% utilization, openEuler has 400\% and 156\% higher schedulability than vanilla Linux and RT-Linux, respectively.


The results of read-write lock is shown in the (e) and (f) of Figure~\ref{fig:sched_re}.
With read-write lock, readers can execute in parallel, while writers has to run sequentially.
The analysis algorithm implemented in SchedCAT is inflation-based analyses~\cite{brandenburg09rw}, which is more pessimistic than the ILP-based approach.
Same as MCS lock, openEuler performs the best over all systems.
Compared to vanilla Linux and RT-Linux, openEuler achieves a maximum improvement of 8$\times$ and 3$\times$, respectively.


\head{Schedulability Imporvement in SUAs.}
To quantify the improvement in schedulability, we measure the schedulability of a scenario using Schedulable Utilization Areas (SUAs).
The SUA of a scenario is the area under its curve in the corresponding schedulability graph. 
Figure \ref{fig:sched_re} shows the improvement in SUA for openEuler in each scenario at the bottom, with the numbers in red and green indicating the SUA improvement of openEuler over Vanilla Linux and RT-Linux, respectively.

\subsection{End-to-End Performance}
\label{sec:endtoend}
In this section, we evaluate the end-to-end response time of a real-time systems in the production environment.
We choose two representative real-time systems, the Core Flight System (cFS)~\cite{cfs_web} and the Robot Operating System 2 (ROS2)~\cite{ros2}, to evaluate the system performance. 
cFS is a widely used middleware for satellite missions.
It can provide common satellite functionalities, such as ground station communication, actuation, and planning. 
We use the OpenSatKit~\cite{opensatkit} to configure the cFS system. 
ROS2 is a popular real-time framework for designing and developing Linux-based robots.
Due to the wide application of ROS2 in robostic systems, the analysis of the real-time performance of ROS2 is an important topic for real-time systems~\cite{ros2_rtss_21, rosm_rtss_22, luo2023modeling,seam_rtss_23,snyc_rtss_23}.
For both frameworks, we bind the real-time processes to the isolated cores, and run the interference workloads on non-isolated cores. 

\head{cFS.} 
We use the flight executive application of cFS to evaluate the end-to-end performance.
The executive is the core of cFS, which is responsible for receiving commands from the ground station, reacting to changes from the simulator, and executing various functions.
Thus the executive requires highly real-time performance. With the executive, the cFS scheduler schedules the HouseKeeping (HK) application, which executes the controling functions of the satellite.


We measure two performance metrics inside cFS: scheduling delay and round-trip communication latency.
The scheduling delay is the time for the cFS scheduler to schedule the HK application.
This delay impacts the system healthness status.
The round-trip communication latency is the time for the HK application to communicate with the sensor.
This latency is safety critical as it determines the system response time of reacting to satellite dynamics.
We set the running period of the HK application to 100 us, and set the communication period between the HK application and the sensor to 200 us.
We use the built-in timer function in cFS to measure latency and record the latency frequency.

\begin{figure*}[!t]

 \begin{minipage}[t]{0.6\linewidth}
    \centering
    \includegraphics[width=\linewidth]{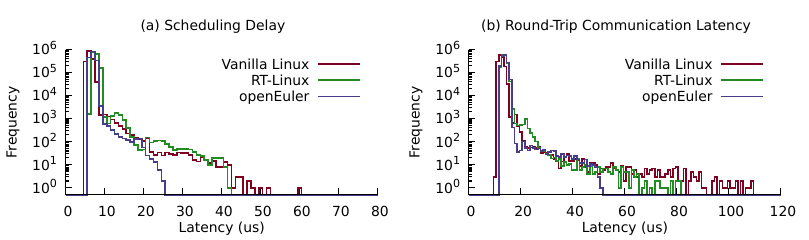}
 \vspace{-1.5em}
  \caption{\small Latency distribution of cFS.}
  \label{fig:cfs_re}
 \end{minipage}%
  \begin{minipage}[t]{0.4\linewidth}
    \centering
    \includegraphics[width=\linewidth]{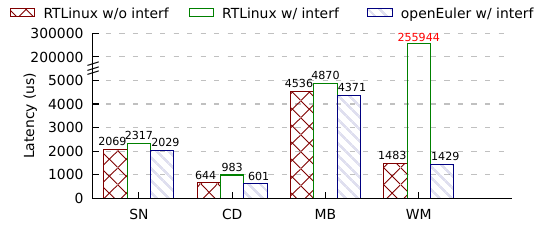}
 \vspace{-1.5em}
	\caption{\small Maximum latency for ROS2.}
    \label{fig:ros_result}
 \end{minipage}%
 \vspace{-2em}
\end{figure*}

\head{Results of cFS.}
Figure~\ref{fig:cfs_re} shows the histogram of the occurrence frequency of cFS scheduling latency  and round-trip communication latency. 
Since existing Linux isolation methods can not eliminate cross-core interference caused by the kernel, both vanilla Linux and RT-Linux suffer substantial interference.
However, openeuler can eliminate the undesired interference and reduce the worst-case latency. For the scheduling delay, the maximum latency is 60 us, 42 us, and 25 us in vanilla
Linux, RT-Linux, and openeuler, respectively.
For the round-trip communication latency, openeuler achieves the lower latency (51 us).
OpenEuler outperforms vanilla Linux (109 us) and RT-Linux (81 us) by 2.1$\times$ and 1.6$\times$, respectively.


\begin{table}[!b]
    \centering
        \caption{\small Scenario Description of the ROS 2 Experiment}

    \begin{tabular}{cc}
        \toprule
        Scenario & Description\\
        \midrule
        SN & 10 nodes, low-throughput, single-executor\\
        CD & 10 nodes, low-throughput, multi-executor\\
        MB & 20 nodes, medium-throughput, single-executor \\
        WM & 20 nodes, high-throughput, multi-executor \\
        \bottomrule
    \end{tabular}
    \label{tab:scenario_description}
    
\end{table}

\head{ROS2.}
A typical ROS application consists of a set of \textit{nodes}.
The scheduling system of ROS 2 is built on the \textit{executor} structure.
Each scheduling thread is an executor that stores all the required communication between nodes and schedules the function calls.
We use the iRobot ROS2 Performance Evaluation Framework \cite{Irobot-Ros} to measures the maximum latency of each communication between two nodes.
The RT priorities of all workload processes on isolated cores are set to 99.
We measured performance under 4 topological structures: {\tt sierra\_nevada} (hereinafter abbreviated as SN, the same below), {\tt cedar} (CD), {\tt mont\_blanc} (MB), and {\tt white\_mountain} (WM).
Table~\ref{tab:scenario_description} describes the basic information for each scenario.
We run each scenario for 5 minutes and record the maximum latency for each type of message transmission.
Considering the poor performance of vanilla Linux in previous experiments, we only compare RTLinux and openEuler.
Meanwhile, we conduct an experiment in RTLinux environment without interference as a baseline to demonstrate the effect of external interference on ROS2.

\head{Results of ROS2.}
Figure~\ref{fig:ros_result} shows the maximum latency of communication under each scenario.
We observed that, the overall impact of interference of multi-executor scenarios (CD, WM) is greater than single-executor scenarios (SN, MB).
Moreover, in the high-throughput scenario (WM), RT-Linux suffered severe packet loss under the influence of interference, reaching a maximum latency of 255ms.
It is fatal for latency-sensitive scenarios.
On the contrary, openEuler significantly reduced the maximum latency to the level of RT-Linux under no interference load.
In most cases, the maximum latency is even lower.
This is because openEuler eliminates most of the disturbances in the system, including those inherently exist in RT-Linux.
In scenarios where no packet loss occurred, openEuler reduced the maximum latency by 1.24$\times$ (2317 us to 2029 us) in single-executor scenarios and 1.64$\times$ (983 us to 601 us) in multi-executor scenarios compared to RT-Linux.

%% file: tex/conc.tex
\section{Conclusions}
\label{s:conc}

OS provides a single execution environment and common management services and is shared by all the cores. 
Thus, the cross-core performance interference caused by the OS itself fundamentally affects the whole system predictability. 
However, current isolation mechanisms are fragmented and fail to guarantee coverage of all interferences.
Existing research lacks both a sufficient view of cross-core interference issues in actual products and systematic design principles for mitigating interference in commercial OSes. 

The paper summarizes the challenges we have encountered in the industry, shares our development experience, and presents our opinions on future system research.
Over the past 6 years, we have fixed various cross-core interference problems in the Linux kernel.
As a result, we significantly reduced the worst-case jitter and achieved improvements of 11.5$\times$ and 3.5$\times$ in the system schedulability over vanilla Linux and RT-Linux, respectively.
For a control loop in satellite software, we achieve a 2.1$\times$ and 1.6$\times$ reduction in worst-case response time compared to vanilla Linux and RT-Linux, respectively.
For nodes communication in the ROS 2 system, we have prevented packet loss caused by interference in RT-Linux and achieved a 1.64$\times$ reduction in maximum latency under generic conditions.
For future real-time OS research, we should systematically identify and mitigate OS interference from three key aspects: task management, resource management, and concurrency management.

%% file: tex/Acknowledgments.tex
\section*{Acknowledgments}
We sincerely thank the RTSS anonymous reviewers for their valuable comments and insightful feedback.
We thank our colleagues at Huawei for their contributions to this work, including but not limited to, Wei Xiong, Wenfeng Zhang, Wenqin Zheng, Guiping Zhang, Wei Du, Xu Wu, HangLiang Lai, and Wenyu Liu. This
work was partly supported by the National Key R\&D of China
(2022YFB4501802), Beijing Natural Science Foundation (L243010) and the National Natural Science Foundation of
China (62141208, 62172009).

Ding Li and Yao Guo are also with Zhongguancun Lab.